\documentclass[12pt]{iopart}
\AtBeginDocument{\RenewCommandCopy\qty\SI}

\usepackage{svg}
\usepackage[nice]{nicefrac}
\expandafter\let\csname equation*\endcsname\relax
\expandafter\let\csname endequation*\endcsname\relax
\usepackage{amsmath}
\usepackage{xfrac}

\usepackage{subcaption}
\usepackage{wrapfig}
\usepackage{sidecap}

\usepackage[
  bookmarks=false,
  colorlinks,
  linkcolor=blue,
  urlcolor=blue,
  citecolor=blue,
  plainpages=false,
  pdfpagelabels,
  final,
  breaklinks=true
]{hyperref}
\hypersetup{
pdftitle={2D quantum-path interference in high-harmonic generation driven by highly-bichromatic fields}, 
pdfauthor={Xiaozhou Zou, Lucie Jurkovičová, Anne Weber, Cong Zhao, Martin Albrecht, Ond\v{r}ej Finke, Alexandr Vendl, Annika Grenfell, Wojciech Szuba, Jaroslav Nejdl, Eric Constant, Margarita Khokhlova, Emilio Pisanty, Ond\v{r}ej Hort and Amelle Zaïr}
}
\usepackage{siunitx}
\usepackage{physics}  
\DeclareSIUnit{\au}{{a.u.}} %%% e.g. $I_0 = \SI{4e14}{W/cm^2}$

\renewcommand{\real}{\mathrm{Re}}

\newcommand{\Ip}{\mathcal{I}_\mathrm{p}}
\newcommand{\degree}{^\circ}

\newcommand{\perc}{R_\mathrm{I}} % intensity ratio
\newcommand{\phasedelay}{\phi_{\omega\text{-}2\omega}} % two-colour relative phase 

\newcommand{\fig}[1]{Fig.\,\ref{#1}}

\newcommand{\eq}[1]{Eq.~\eqref{#1}}

\renewcommand{\vec}[1]{\mathbf{#1}}
\newcommand{\im}{\mathrm{i}} % imaginary unit
\renewcommand{\e}{\mathrm{e}} % e

\usepackage{amssymb}
\usepackage[utf8]{inputenc}
\usepackage{printlen}

\usepackage{graphicx}
\usepackage[sorting=none,citestyle=numeric-comp]{biblatex}
\definecolor{Cerulean}{rgb}{0.0, 0.48, 0.65}

\providecommand{\keywords}[1]
{
  \small	
  \textbf{Keywords:} #1
}

\begin{document}
\title[2D QPI in HHG driven by highly-bichromatic laser fields] {2D quantum-path interference %quantum-path interference
in high-harmonic generation driven by highly-bichromatic fields}

\author{Xiaozhou Zou$^{1,\dag}$, Lucie Jurkovi\v{c}ov\'{a}$^{2,3,\dag}$, Anne Weber$^{1,\dag}$, Cong Zhao$^{1}$, Martin Albrecht$^{2,3}$, Ond\v{r}ej Finke$^{2,3}$, Alexandr Vendl$^{2}$, Annika Grenfell$^{2}$, Wojciech Szuba$^{2}$, Jaroslav Nejdl$^{2,3}$, Eric Constant$^{4}$, Margarita Khokhlova$^{1}$, Emilio Pisanty$^{1}$, Ond\v{r}ej Hort$^{2}$ and Amelle Za\"{i}r$^{1}$}
\address{$^1$ Attosecond Quantum Physics Laboratory, King’s College London, WC2R 2LS, UK}
\address{$^2$ ELI Beamlines Facility, The Extreme Light Infrastructure ERIC, Za Radnic\'{i} 835, Doln\'{i} B\v{r}e\v{z}any 252 41, Czech Republic}
\address{$^3$ Czech Technical University in Prague, FNSPE, B\v{r}ehov\'{a} 7, Prague 1 115 19, Czech Republic}
\address{$^4$ Universit\'{e} Claude Bernard Lyon 1, CNRS, Institut Lumière Matière (ILM), 69622 Villeurbanne France}
\ead{xiaozhou.1.zou@kcl.ac.uk\\
     \qquad \;\; amelle.zair@kcl.ac.uk}
     
\vspace{10pt}
\begin{indented}
\item[]{\dag} equal contribution.
\end{indented}

\begin{abstract}
We experimentally observe a new type of quantum-path interference, in two-dimensions (2D-QPI), in high-harmonic generation (HHG) driven by an orthogonally-polarised highly-bichromatic field. 
This regime is marked by comparable intensities of the two orthogonal colours. 
In this highly-bichromatic regime, we demonstrate that 2D-QPI 
is encoded in the measured harmonic intensity modulations with respect to the relative phase of the two-colour field. 
The modulations of the odd-order harmonics show a monomodal behaviour, whereas the even harmonics are modulated in a bimodal structure. 
Our calculations using the strong-field approximation and saddle-point method disentangle contributions from multiple quantum orbits in this HHG regime, revealing that the dipole response for both odd and even harmonics inherits the dynamic symmetry of the orthogonally-polarised driving field. This new type of 2D-QPI offers a novel route to 
HHG spectroscopy of attosecond electron dynamics by lifting up the dimensionality of the quantum paths involved in the interference.

\end{abstract}
\keywords{laser-matter interaction,  high-harmonic generation, quantum-path interference}

\section{Introduction}
High-harmonic generation (HHG) is a pivotal process in strong-field light-matter interactions, leading to the production of attosecond pulses in the extreme ultraviolet (XUV) regime \cite{paul2001observation, hentschel2001attosecond, khokhlova2023shining}.
The HHG process can be described in terms of the three-step model \cite{corkum1993plasma, kulander1993dynamics}, which highlights that the generated attosecond XUV pulses are highly influenced by the subcycle temporal structure of the driving laser field.
Thus, a common technique to tailor the resulting XUV pulses is to employ bichromatic fields to drive HHG.
Mixing a fundamental ($\omega$) field with an additional commensurable frequency component, with control over the components' frequencies \cite{schiessl2006enhancement}, polarisation \cite{habibovic2022high} and the two-colour relative phase \cite{kim2005highly} and intensity ratio \cite{raab2024highly}, allows to harness these subcycle dynamics.
In these previous investigations the weak second-harmonic ($2\omega$) field is often considered as a perturbative probe.

As one of these bichromatic configurations, HHG in orthogonally polarised two-colour fields has been intensively investigated \cite{kim2005highly, kim2005generation, shafir2012resolving, niikura2011probing, shafir2009atomic, lambert2015towards, brugnera2011trajectory}. 
One notable application was the enhancement of the total harmonic intensity \cite{kim2005generation,roscam2021divergence} compared to HHG driven by a single-colour field. 
Because electron dynamics are no longer limited to 1D quantum paths but are instead unfolded into a 2D plane, high-harmonic spectroscopy techniques can access quantum dynamics in systems where control over the recollision angle would allow tomographic probing. Thus, it has been demonstrated that it is possible to retrieve the ionisation and recombination times of electronic wavepacket trajectories that lead to high harmonic XUV emission in atoms \cite{shafir2012resolving,zhao2013determination}, probe valence electron wavepackets \cite{niikura2011probing} and atomic wavefunctions \cite{shafir2009atomic}. The retrieval scheme is generally based on utilising harmonic intensity modulation to extract the microscopic electron dynamics. 

It is only recently that highly-bichromatic laser drivers have attracted interest in HHG, where the additional second-harmonic field is no longer a perturbation \cite{lambert2015towards,brugnera2011trajectory}. In this regime, the second-harmonic field can be used as a tool to significantly reshape the ionisation time and steer electronic wavepacket trajectories in the continuum, which results in more than one relevant recombination event in each half-cycle.
It is also expected to fully make use of the 2D steering of the electronic wavepacket in the continuum, to provide interesting angles of recollision and probe systems tomographically. 
Several studies in similar strong $2\omega$ regimes have reported control over the HHG process, showing that the polarisation of harmonics can reach high degrees of ellipticity \cite{lambert2015towards, baykusheva2018chiral}, and allowing selection over quantum trajectories \cite{brugnera2011trajectory}.

Full realisation of this control over the emitted high harmonics requires the experimental manipulation of the driving field to be supported by a deep theoretical understanding of the dynamics.
For HHG, theory typically takes the form of simulations of the time-dependent Schrödinger equation (TDSE), which can be matched to intuitive classical-trajectory dynamics \cite{corkum1993plasma, kulander1993dynamics} through the quantum-orbit formalism of the strong-field approximation (SFA) \cite{lewenstein1994theory,salieres2001feynman}.
This provides a quantitative model based on saddle-point methods %(SPM) 
that directly matches the classical intuition.
For bichromatic fields, both TDSE simulations \cite{murakami2013high, frolov2016control} and classical models \cite{he2010interference} have been reported.
However, to the best of our knowledge, the key link provided by the quantum-orbit description has only been applied to bicircular fields with inherent dynamical symmetry \cite{milosevic2019quantum,milosevic2000generation,pisanty2014spin} and to perturbative two-colour configurations \cite{zhao2013determination}. 
At present, though, a quantum-orbit description for HHG driven by bichromatic fields with comparable intensities of the two colours has not yet been clearly established, and poses significant and novel challenges of its own.

%%%
In this work, we experimentally demonstrate a new type of quantum-path interference (QPI) \cite{zair2008quantum} using a highly-bichromatic field with orthogonal components. 
For the first time, we apply the saddle-point methods %SPM 
to HHG in this regime, which allows us to distinguish between the contributions from various 2D trajectories involved in the physical process. 
The observed bimodal even-harmonic intensity modulations with respect to the two-colour relative phase are identified as a 2D-QPI feature in the highly-bichromatic regime, where QPI truly unlocks the potential of the extra control handles provided by the additional field by allowing drastic reshaping of the sub-cycle dynamics through their strong dependence on the %inter-colour
relative two-colour phase. 
Ultimately, the further analysis of the quantum orbits shows how the dynamical symmetry of the driving field is imprinted onto the harmonic dipole lifting up the dimensionality of the QPI.

\section{Experimental observations}
\subsection{Experimental schematic}
\begin{figure}[ht]
\centering
\includegraphics[width=0.5\textwidth]{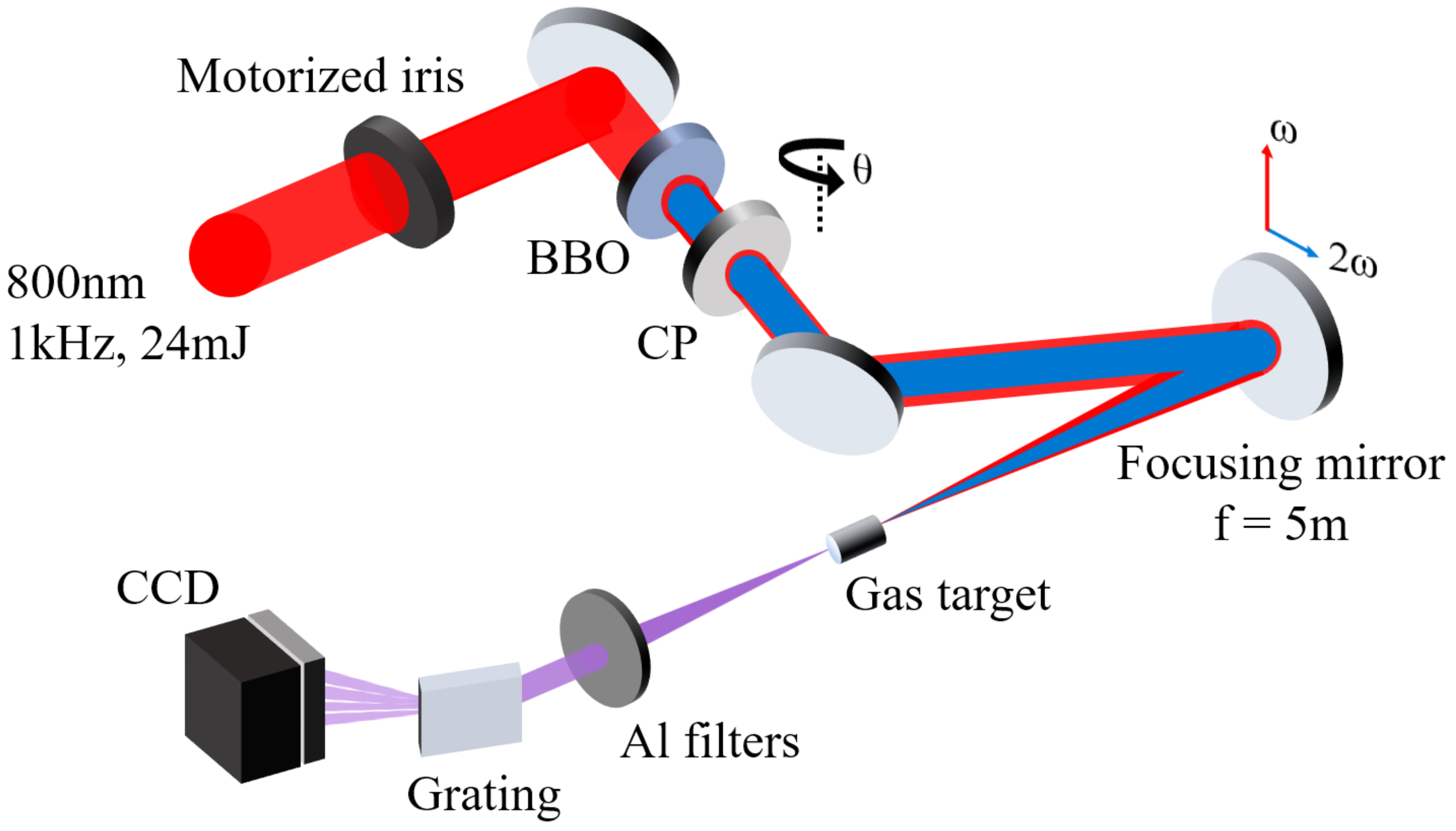}
\caption{Schematic of the experimental setup. The fundamental beam generates the second harmonic in a BBO crystal (type I). The two-colour relative phase is controlled by the fine rotation of the calcite plate (CP) $\theta$ angle. The bichromatic field is focused into an argon gas cell by a spherical mirror with a 5 m focal length, and the high harmonics are characterised by a flat-field spectrometer.}

\label{setup}
\end{figure}

We performed the experiment using the High-Harmonic Beamline at the Extreme Light Infrastructure (ELI Beamlines Centre, ELI ERIC) \cite{hort2019high}; a schematic of the setup is shown in \fig{setup}.
The main infrared driver ($\omega$) for HHG is the L1 Allegra laser system ($\SI{15}{fs}$, $\SI{800}{nm}$, $\SI{24}{mJ}$, $\SI{1}{kHz}$) based on optical parametric chirped pulse amplification (OPCPA) \cite{antipenkov2023l1}. The power of the input pulse is controlled using a motorised iris. A $\beta$-barium borate (BBO) crystal ($\SI{200}{\mu m}$, type-I, cut at $29.2\degree$) is used for second-harmonic ($2\omega$) generation with polarisation orthogonal to the $\omega$ driver. 
A calcite plate ($\SI{500}{\mu m}$, cut at $45\degree$) is mounted on a motorised rotational stage allowing precise control over the orientation of the optical axis, which enables fine scanning of the two-colour relative phase $\phasedelay$. 

The collinearly propagating fundamental ($\omega$) and second-harmonic ($2\omega$) beams are focused by a spherical mirror with a focal length of \SI{5}{m} (f-number of 266, corresponding to beam diameter of \SI{18.8}{mm} limited by the iris diameter) onto a continuous argon gas cell (\SI{0.5}{mm}, \SI{40}{mbar}) to drive HHG. 
The focal spot was characterised by imaging it through a lens positioned behind the focus. The measured value of the IR beam diameter in the focus for the iris diameter
\SI{18.8}{\mm} was \SI{280}{\micro\meter}.
The energy of the two-colour driving beam was measured using an energy meter in combination with a dichroic mirror positioned in the beam path after the focusing optics. 
This configuration enabled simultaneous measurement of both fundamental and second-harmonic beams, the $2\omega$ conversion efficiency is 11\%.
The intensities of the $\omega$ and $2\omega$ laser pulses at the focus are estimated to be 
$I_\omega \approx \SI{1.5e14}{W/cm^2}$ and $I_{2\omega}\approx  \SI{1.8e13}{W/cm^2}$, respectively, giving the intensity ratio $\perc = I_{2\omega}/I_\omega \approx 0.12$.
Two aluminium filters ($\SI{1}{\mu m}$ and $\SI{0.2}{\mu m}$) are placed \SI{6}{m} after the gas cell to block the bichromatic driving laser and low-order harmonics. 
The emission of high-order harmonics is detected using a flat-field XUV spectrometer composed of a blazed concave grating and a soft-X-ray CCD camera (Andor) placed \SI{9.2}{m} after the gas target.
\subsection{Experimental results}

The HHG spectrum in the far field varies with two-colour relative phase $\phasedelay$ (defined in Eq.~\eqref{E-definition} below). In \fig{selection}, HHG spectra for two calcite plate $\theta$ angles are shown; the angle difference corresponds to approximately $\pi/2$ phase difference between the two panels (1/4 cycle of the $\omega$ field). We show in \fig{selection} (a) an HHG spectrum with a larger divergence, associated with long trajectories contributing to the off-axis signal at $\phasedelay= 0$, corresponding to a figure-of-eight Lissajous figure for the electric field, through the theoretical modelling described below.
In comparison, the harmonics are less divergent for $\phasedelay= \pi/2$ (\fig{selection} (b)), corresponding to a crescent-shaped Lissajous figure for the field, and the overall intensity yield of HHG is considerably higher which is consistent with previously reported work \cite{kim2005highly, brugnera2011trajectory}.

\begin{figure}
    \centering
    \includegraphics[width=0.75\linewidth]{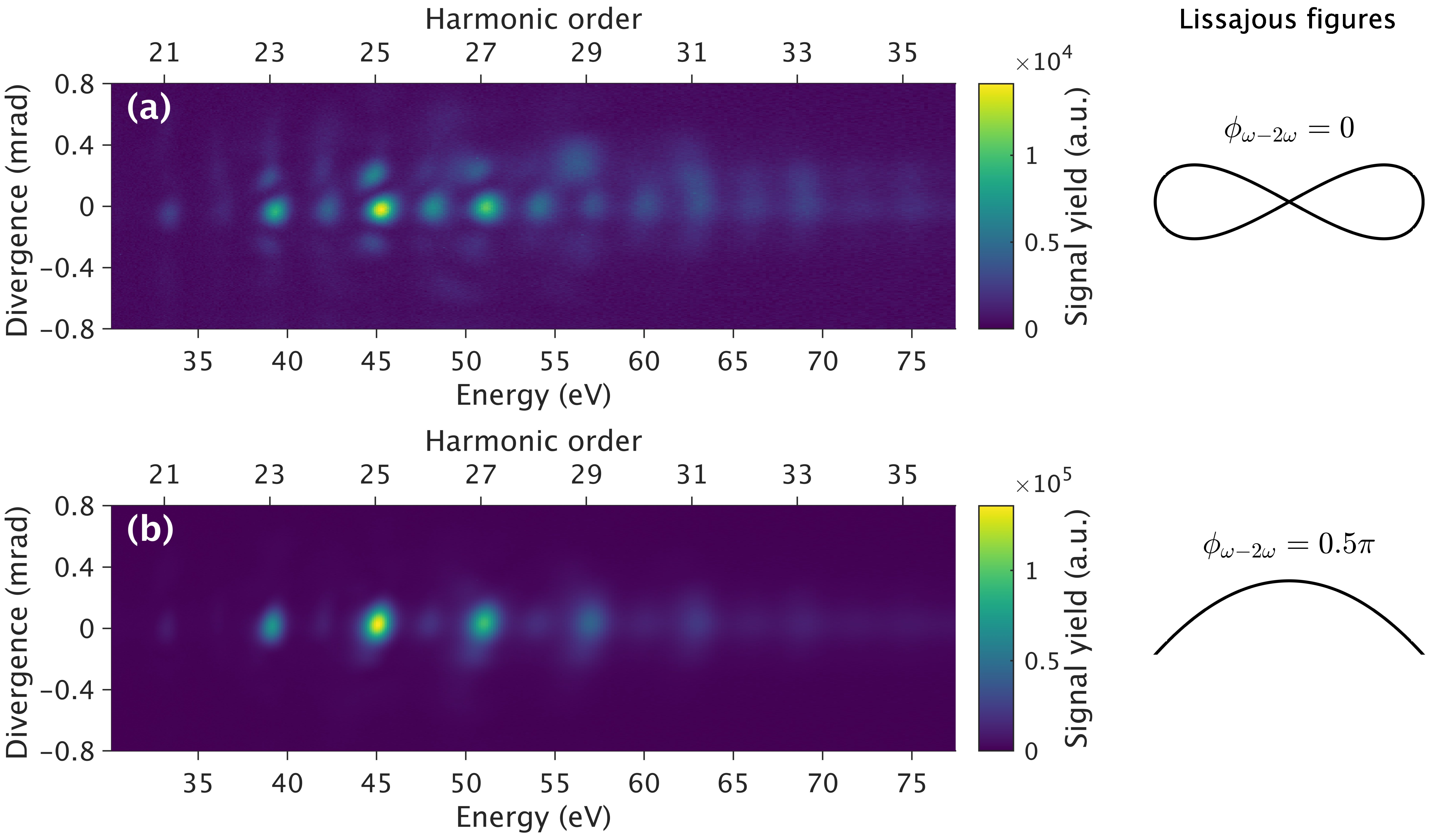}
    \caption{Far-field spatially-resolved HHG spectra for relative phase of (a)~$\phasedelay=0$ and (b)~$\phasedelay=\pi/2$, corresponding Lissajous figures are illustrated on the right.}
    \label{selection}
\end{figure}
%%% delay scan

\begin{figure}[htbp]
\centering
\includegraphics[width=0.95\linewidth]{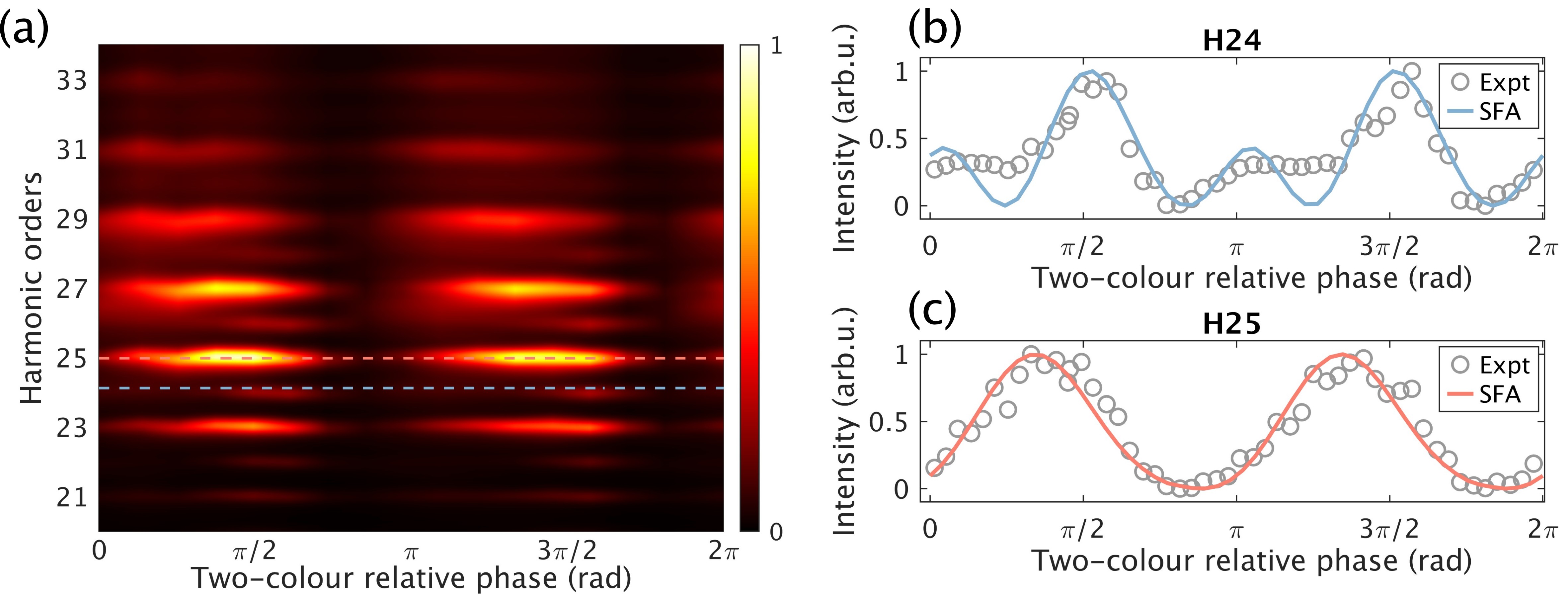}
\caption{(a)~Two-colour relative phase scan; (b,c)~experimental data (circles) and SFA simulation (lines) of total intensity yield modulation for harmonic order (b)~24 and (c)~25.}
\label{fig:delay scan}
\end{figure}

The measured HHG spectra as a function of $\phasedelay$ are shown in \fig{fig:delay scan}(a), the spectrum is integrated along the spatial divergence axis shown in \fig{selection}. Individual intensity modulations are shown in \fig{fig:delay scan} (right) for an even (H24, b) and an odd (H25, c) harmonic order, where dots represent the experimentally measured data, with $\phasedelay$ wrapped in $2 \pi$. The lines are the results from the SFA calculation (see Theoretical calculation section).
The simulated intensity modulations are first fitted to a two-term Fourier model for harmonic order $q$: $f_{\mathrm{fit}}^{\mathrm{(q)}}(t)= a_0^{(q)} + a_1^{(q)}\cos(\omega t) + b_1^{(q)}\sin(\omega t) + a_2^{(q)}\cos(2\omega t) + b_2^{(q)}\sin(2\omega t)$ via the least-squares algorithm. The absolute phase difference is not directly measured in the experiment, so the measured intensity modulation is shifted to match the SFA calculation; the shift~$\tau$ is obtained by fitting the SFA result $f_{\mathrm{fit}}^{\mathrm{(q)}}(t-\tau)$ to the measured data using the least-squares algorithm.

For each harmonic order, a scan over the relative phase yields an oscillation in harmonic intensity. 
These oscillations exhibit distinct structures between even and odd harmonic orders, as illustrated in \fig{fig:delay scan}:
for odd orders (c) we find a monomodal oscillation, while for even orders (b), the oscillation demonstrates a bimodal structure. Additional intensity modulations for other orders are provided in the supplementary material. The origin of such features can be explained using the SFA-based quantum-orbit model that allows us to express the harmonic response in terms of different trajectories' interference.

\section{Theoretical calculations}
\subsection{Quantum orbits in the SFA framework}

%%% Driving field definition
We calculate the HHG signal generated by our highly bichromatic field, for which the electric field is defined as
\begin{equation}
\vec{E}(t)= {E_\omega} \sin(\omega t) \hat{\vec{x}} + {E_{2\omega}} \sin(2\omega t + \phasedelay) \hat{\vec{y}}\,
\label{E-definition}
\end{equation}
with two-colour relative phase $\phasedelay$.
We define the intensity ratio $\perc$ via the ratio of amplitudes of the two field components: $E_{2\omega}/E_\omega = \sqrt{\perc}$. 
%%% Dipole calculation, Saddle point selection
To simulate the experimentally observed HHG signal, we calculate the microscopic HHG response in terms of the dipole of the harmonic radiation $\vec{D}(q \omega)$, for a given harmonic order $q$.
The intensity of a given harmonic is then given by $I(q \omega) = \frac{(q \omega)^4}{2 \pi c^3} |\vec{D}(q \omega)|^2 $ \cite{nayak2019saddle}.
The theoretical calculation of the microscopic HHG response has been performed using the SFA to simulate the harmonics' intensity modulation in the experiment. 
We use the SFA \cite{nayak2019saddle, smirnova2014multielectron} such that the dipole of the harmonic radiation for harmonic order $q$ can be expressed in terms of the integral over ionisation ($t_i$) and recombination ($t_r$) time as
\begin{equation}
      \vec{D}(q \omega) =  
        \int_{-\infty}^{\infty} \,\mathrm{d}t_r
        \int_{-\infty}^{t_r} \,\mathrm{d}t_i ~
        \vec{d}\left(\vec{p}_s + \vec{A}(t_r)\right)
        \Upsilon \left(\vec{p}_s, t_i \right) 
        \left(\frac{2 \pi}{\im (t_r - t_i)} \right)^{\sfrac{3}{2}}
      \e^{\im S(\vec{p}_s,t_i,t_r)} + \text {c.c.}
\label{Dipole integral}
\end{equation}
with the dipole transition matrix element $\vec{d}(\vec{k}) = \frac{\im \sqrt{2} \vec{k} }{\pi \sqrt{2 \Ip} (\vec{k}^2 + \sqrt{2 \Ip} )^2 }$,%
\footnote{
We model the spherically-symmetric argon gas using a hydrogenic orbital, as per usual practice \cite{Le2016}. While there are some difference between $s$-shell and $p$-shell atoms \cite{Pisanty2017}, these are expected to be minimal in this context.
}
ionisation amplitude $\Upsilon(\vec{k_i}) = \frac{1}{ 2 \pi \sqrt{\Ip}}$
and the stationary momentum $\vec{p}_s = \vec{p}_s(t_i, t_r) = -\frac{1}{t_r - t_i} \int_{t_i}^{t_r} \vec{A}(t)\,\mathrm{d}t$; we use atomic units ($\mathrm{a.u.}$) unless specified otherwise.
The exponent in \eq{Dipole integral} contains the 
semi-classical action
\begin{equation}
        S(\vec{p}_s,t_i, t_r) = -\int_{t_i}^{t_r} \mathrm{d}t\,
        \left[ \Ip + \frac{1}{2} \left( \vec{p}_s  + \vec{A}(t)\right)^2 \right]
        + q \omega t_r \,
    \end{equation}
As the integral in~\eq{Dipole integral} is highly oscillatory, it is convenient to solve it using the saddle-point methods, which reduces the integration to a sum of contributions stemming from saddle points of the exponent \cite{nayak2019saddle, smirnova2014multielectron}:
 \begin{align}
      \vec{D} (q \omega) &=
        \sum_s \vec{D}_s(q \omega) \\
      &= \sum_s \frac{2 \pi}{\sqrt{-\mathrm{det}(S'')}} 
        \vec{d}\left(\vec{p}_s + \vec{A}(t_{r,s})\right)
        \Upsilon \left(\vec{p}_s, t_{i,s} \right) 
        \left( \frac{2 \pi}{\im (t_{r,s} - t_{i,s})} \right)^{\sfrac{3}{2}}
      \e^{\im S(\vec{p}_s, t_{i,s} ,t_{r,s})} + \text {c.c.}\,
      \label{eq:dipolesum}
\end{align}
Here $S^{''}$ is the Hessian matrix of total action $S(t_i, t_r)$ and the saddle points $(t_{i,s},t_{r,s})$ can be found by solving the system of equations
\begin{align}
\begin{split}
    \tfrac{\partial S}{\partial t_r} &= \tfrac{1}{2} \left( \vec{p}_s + \vec{A}(t_r) \right)^2 + \Ip -  q \omega = 0 \\
    \tfrac{\partial S}{\partial t_i} &= \tfrac{1}{2} \left( \vec{p}_s + \vec{A}(t_i) \right)^2 + \Ip = 0 \,
\end{split}
\end{align}
The summation in~\eq{eq:dipolesum} then includes all saddle points $s$ that are part of a steepest-descent integration domain. 
Each saddle point corresponds to an electron trajectory defined by its ionisation and recombination time, and hence gives rise to the understanding of the HHG process in terms of several quantum orbits \cite{milosevic2000generation,salieres2001feynman}.

When using a monochromatic driving laser field, the trajectories within one half-cycle of the field can easily be classified as `short' and `long' trajectories with regards to their travel time $ \real{\left(t_r - t_i \right)}$. 
As we add a strong $2\omega$ field to the fundamental driver, there are usually more than two saddle points (trajectories) within one half-cycle. 
Furthermore, the multi-index classification scheme used in e.g. \cite{milosevic2019x, popruzhenko2002laser, milosevic2002role, chipperfield2005conditions} breaks down as the number of ionisation bursts changes with relative intensity $\perc$ and relative phase $\phasedelay$.
A generic classification of all those saddle-point solutions is therefore far from trivial and requires to account for coalescences and branch cuts of the solutions \cite{pisanty2020the,raz2012spectral}; indeed, the different trajectories can be seen to form simply different branches of a single unified Riemann surface \cite{pisanty2020the}. 
As a general rule, a rigorous method to decide on whether the contribution from a chosen saddle point should be included into the final harmonic dipole has long been recognised, if somewhat quietly, as a challenging or even inaccessible problem \cite{milosevic2000generation, milosevic2022negative}.
However, determining the subset of saddle points that lie on a valid steepest-descent integration domain is independent of their classification: 
a given saddle point is relevant if its attached steepest-ascent manifold intersects the original integration domain \cite{weber2025universalapproachsaddlepointmethods}. This method allows us to consistently evaluate the dipole with accounting all the relevant saddle points in our configurations.

%%% Example: saddle point times

\begin{figure}[htbp]
\centering
\includegraphics[width=0.6\textwidth]{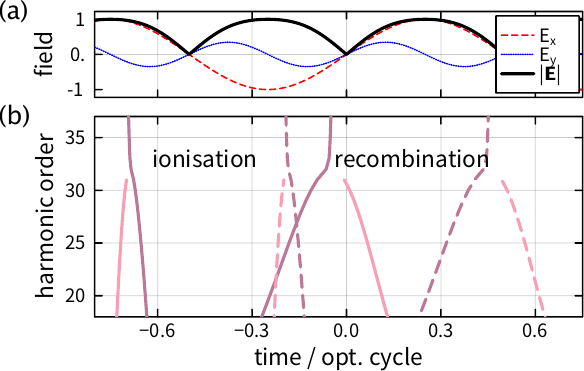}
\caption{Ionisation and recombination times for a two-colour field with intensity ratio $\perc = 12\%$ and $\phasedelay = 0$.
(a)~The total amplitude (black solid line) and its component fields~-- the fundamental component polarised in the $x$ direction (red dashed line) and the $2\omega$ component polarised in the $y$ direction (blue dotted line).
(b)~Real parts of the saddle-point times (ionisation on the left-hand side, recombination on the right-hand side) across a range of harmonic orders.
For the shown field configuration, there is one short (purple) and one long (pink) trajectory per half-cycle; the first half-cycle solutions are drawn in solid lines, the second half-cycle ones in dashed lines.}
\label{saddlepoints}
\end{figure}
Within this work, we restrict the intensity ratio $\perc$ to $12\%$ as used in the experiment.
For the two-colour relative phase $\phasedelay = 0$, we show in \fig{saddlepoints} the most relevant ionisation and recombination times for a range of harmonic orders ((b), bottom panel), alongside the respective electric field of the driving laser ((a), top panel).
We can clearly identify the familiar structure of a pair of short (purple) and long (pink) trajectories per half-cycle, where the short trajectories need to be discarded beyond the high-order harmonic cutoff at H29. 
We have drawn the first half-cycle solutions with solid lines and the second half-cycles' solutions with dashed lines, for consistency with the subsequent figures.
Throughout the scan of the relative phase $\phasedelay$ we can distinguish the long trajectories from their shorter counterparts, and for the majority of the scan we can identify the familiar structure of two trajectories per half-cycle as shown in \fig{saddlepoints}. 

\subsection{2D quantum-path interference in the two-colour relative phase scan}

%%% "Fig 5" Major polarisation axes
\begin{figure}[t!]
\centering\includegraphics[width=\textwidth]{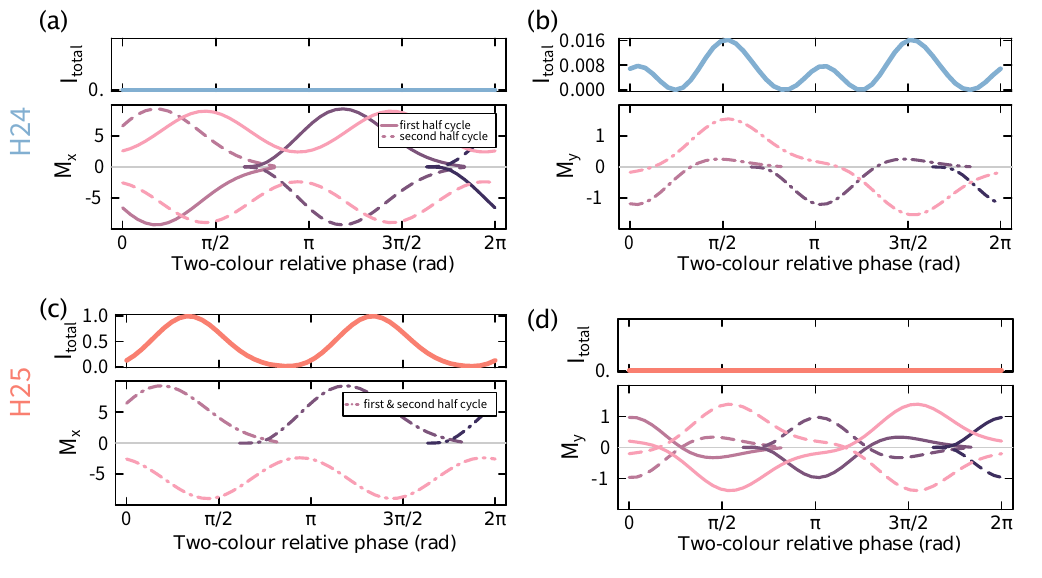}
\caption{Total harmonic intensity modulation (top subpanels) as well as the main contributions from the most dominant trajectories (bottom subpanels) throughout a scan of the relative phase $\phasedelay$. Those contributions are given in terms of the components of the major polarisation axes of the harmonic field ellipse, $M_x$ and $M_y$. 
In the top row, we show (a) the $x$ (parallel to the $\omega$ polarisation) and (b) the $y$ (parallel to the $2\omega$ polarisation) components for H24, and in the bottom row (c,d) the same components for H25.}
\label{fig:polaxes}
\end{figure}

We now turn to the evaluation of the harmonic signal throughout a scan of the relative phase $\phasedelay$ as explained above.
To explain the origin of the observed intensity oscillations, in \fig{fig:polaxes} we plot the harmonic response for two subsequent harmonic orders (H24 in (a,b), H25 in (c,d)) and look at the $x$ (a,c) and $y$ (b,d) components separately.
Within each panel, we show the total harmonic signal (upper panel) and the contribution from the various trajectories (lower panel).
The components of the total intensity are calculated as 
$I_{i,\mathrm{total}} \propto (q \omega)^4 |\vec{D}_i(q \omega)|^2 $ (for $i = x,y$) and normalised to the maximum of the oscillation shown in panel (c).
% $I(q \omega) = \frac{(q \omega)^4}{2 \pi c^3} |\vec{D}(q \omega)|^2 $.

\begin{wrapfigure}{r}{0.35\linewidth}
  \begin{center}
    \includegraphics[width=0.7\linewidth]{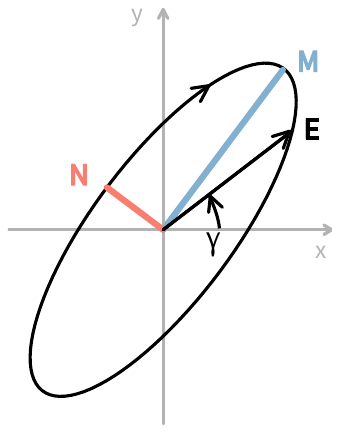}%% linewidth is as a percentage of the box
  \end{center}
  \caption{Polarisation ellipse.}
  \label{fig:ellipse}
\end{wrapfigure}

The contributions of the various trajectories are then shown in terms of their major polarisation axes $\vec{M}$ of the harmonic dipole ellipse.
That is, the electric field associated with a harmonic dipole 
$\vec{D}_s(q \omega)$ can be expressed as
\begin{align}
  \begin{aligned}
    \vec{E}_{q\omega}(t) = \vec{D}_s(q \omega)\, \e^{\im q \omega t}
    = \e^{\im \gamma} \left(\vec{M} + \im \vec{N}\right)\, ,
  \end{aligned}
  \label{eq:ellipse}
\end{align}
where $\vec{M}$ and $\vec{N}$ are the major and minor polarisation axes, respectively, and $\gamma$ is known as a rectifying phase \cite{berry2004index}, 
as shown in \fig{fig:ellipse}, the elliptical field-shape representation of~\eq{eq:ellipse}.

For the harmonic signal in \fig{fig:polaxes} we only show the $x$ and $y$ components of $\vec{M}$, as the intensity along the minor polarisation axis $\vec{N}$ is negligible, with $|\vec{N}|/|\vec{M}|$ in the order of $5\%$. From \fig{fig:polaxes} we can now understand the following:
for H24 (top row) the trajectories from within the two subsequent half-cycles (solid and dashed lines respectively) produce dipoles with opposite orientation, and their polarisation axes are symmetric along the $x$ axis (see bottom figure in panel (a)). 
As a result, their contributions interfere destructively in the $x$ direction and the $x$ component of the total intensity vanishes (panel (a)).
The $y$ components of the trajectories, however, are oriented in the same direction for the two half-cycles (dash-dotted lines in the bottom figure of panel (b)).
Hence, their contributions interfere constructively to produce an oscillating, non-zero harmonic signal along the $y$ direction (panel (b)).
For H25 the situation is inverse:
The $x$ components of the individual trajectory's major dipole axes are oriented in the same direction for the two half-cycles and interfere constructively. 
This produces an oscillatory total signal (panel (c)). 
Conversely, the $y$ components have opposite orientation for the two half-cycles and therefore interfere destructively such that the total signal is zero (panel (d)). 

This correspondence between the polarisation of the harmonics and their order is rooted in the symmetry of the field \cite{alon1998selection}, and has been used as a proxy to infer polarisation characteristics in the weakly-bichromatic regime \cite{shafir2012resolving,zhao2013determination}.
In our case, the resulting oscillations differ significantly between the $x$ and $y$ components for the various trajectories, and this assignment allows us a deeper insight into the measured mono- and bi-modal oscillations in the signal.
The oscillation pattern along the two axes can be related to the frequencies of the driving field: the observed modulation of the harmonic signal along the $x$ axis follows the $\omega$ field, and the oscillation along $y$ follows the $2 \omega$ field. 

%%% trajectories / displacements
\begin{figure}[htbp]
\centering
\includegraphics[width=\linewidth]{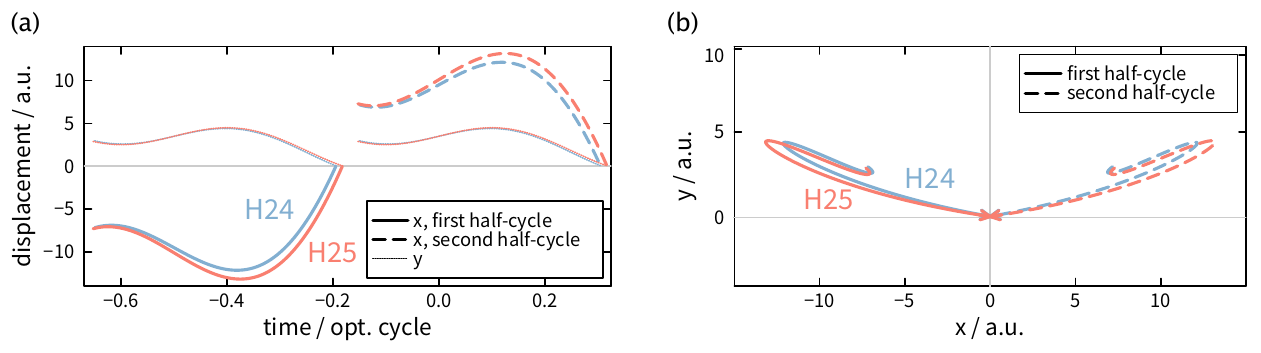}
\caption{Displacements $\vec{s}(\real(t))$ for two short trajectories within one optical cycle for a two-colour field as above, with $\phasedelay=0$, for the two subsequent harmonic orders $24$ and $25$ shown (a)~over time, and (b)~in space.}
\label{fig:displacements}

\end{figure}
To visualise this further, in \fig{fig:displacements} we show the electron's displacement, i.e.\ the orbit during the propagation in the laser field, which is given by
\begin{equation}
      \vec{s}(t) = \real \left[\int_{t_{i,s}}^{t} \left( \vec{p} + \vec{A}(t') \right) \,\mathrm{d}t' \right] \,
      \label{eq:trajectory}
\end{equation}
For the two most dominant trajectories for harmonic order 24 (red) and 25 (blue) generated from a field with $\phasedelay = 0$.
That is, these orbits correspond to the two short trajectories whose dipole polarisation axes are depicted in \fig{fig:polaxes}(a) and (b) in violet.
In \fig{fig:displacements}(a) we show the components of the displacement in $x$ and $y$ direction. 
We find that the $x$ component is opposite for the two half-cycles, i.e.\ negative for $t \lesssim {-20}$ (solid lines) and positive for $t \gtrsim -20$ (dashed lines), while the $y$ direction is identical in both half-cycles (thin dotted lines). 
In \fig{fig:displacements}(b) we plot exactly the same orbits, but in $x$-$y$ space. 
As expected, we find the trajectories start away from the core, where the electron appears in the continuum, are subsequently driven further away and then return in a slingshot-movement back to the core (at $(0,0)$) where they recombine.
Most importantly, we can clearly see that the trajectories in the first (solid line) and second (dashed line) half-cycles appear symmetric with respect to the $x$ axis. 

Generally, while the trajectories only change slightly from H24 to H25, \fig{fig:displacements}(a) and (b) both clearly demonstrate the symmetry imposed by the driving field. 
That is, the trajectories in the second half-cycle are flipped along the $x$ axis with respect to the trajectory in the first half-cycle. 
Moreover, both figures show that the displacements in $y$ direction (up to $\approx 5\, \mathrm{a.u.}$) are smaller than those in $x$ direction (up to $\approx \pm 13\, \mathrm{a.u.}$). 
This can readily be explained as the $y$ component of the driving field (which oscillates with $2\omega$) is weaker (reminder: intensity ratio $\perc = 12\%$) than the $x$ component.

\section{Conclusion}
We demonstrate that 2D-QPI %a new type of QPI
can be observed in HHG from gas-phase atoms driven by highly-bichromatic fields. The interference leads to different intensity modulation patterns for the odd and even harmonics, monomodal and bimodal, respectively, over a scan of the two-colour relative phase between the $\omega$ and $2\omega$ field components.

We performed calculations within the SFA framework, which are able to reproduce and give further insight into those intensity modulations.
For even harmonic orders the component of the dipoles along the $\omega$ polarisation ($x$) cancel each other out and result in a vanishing signal in the $x$ direction, while along the $2\omega$ polarisation in the $y$ direction they add up constructively; 
for odd harmonic orders, the $x$ components add up and the $y$ components interfere destructively. 
These dynamical symmetries of the field can immediately be visualised by the electron displacements of the short trajectories within each half-cycle.
Those trajectories in the second half-cycle appear flipped along the $x$ axes, while the $y$ component remains the same.

The bimodal nature of the even-harmonic intensity modulations can thus be directly traced to the interference between the $y$ components of the trajectories' dipoles.
This inherits a more complex behaviour from the higher frequency of the field in that direction, and it is explicitly revealed in the spectral domain since the symmetry of the field requires the even harmonics to be polarised in the same direction as the $2\omega$ driver.

Our results demonstrate a clear signature, and give a comprehensive picture, of 2D-QPI produced in HHG driven by highly-bichromatic laser fields. 
This introduces the bichromatic phase delay as a new observable for QPI, applicable to both co-linear (1D) and orthogonally polarized (2D) two-colour driving field geometries.
Highly-bichromatic fields are an emerging and increasingly available and powerful tool to shape attosecond HHG light, with significant flexibility coming from the large number of handles they provide. This generation of attosecond light sources forms the first step towards 
new multidimensional attosecond spectroscopy techniques to resolve ultrafast electronic dynamics and tomographic information.
The understanding of QPI provided by our results builds the foundation for detailed control over harmonic emission in this regime.

\section*{Acknowledgments}
This work has been supported by Royal Society funding via A.Z.'s research project ‘XAWO’ IES\textbackslash{}R3\textbackslash{}203022; A.Z. acknowledges the ELI-ERIC beam time allocation at ELI-beamlines; A.W., M.K.\ and E.P.\ acknowledge Royal Society funding under URF\textbackslash{}R1\textbackslash{}211390 and URF\textbackslash{}R1\textbackslash{}231460. 
\printbibliography
\end{document}